\documentclass[12pt,preprint]{aastex}
\usepackage{apjfonts}
\usepackage{colordvi}

\begin{document}
\title{Coronal Thick Target Hard X Ray Emissions and Radio Emissions}
\author{Jeongwoo Lee\altaffilmark{1,2}, Daye Lim\altaffilmark{2},
G. S. Choe\altaffilmark{2}, Kap-Sung Kim\altaffilmark{2}, and
Minhwan Jang\altaffilmark{3}} \affil{1. Physics Department, New
Jersey Institute of Technology, Newark, NJ 07102} \affil{2. School
of Space Research, Kyung Hee University, Yongin 446-701, Korea}
\affil{3. Department of Astronomy and Space Science, Kyung Hee
University, Yongin 446-701, Korea}

\begin{abstract}
Recently a distinctive class of hard X ray (HXR) sources located in
the corona was found, which implies that the collisionally thick
target model (CTTM) applies even to the corona. We investigated
whether this idea can independently be verified by microwave
radiations that have been known as the best companion to HXRs. The
study is made for the GOES M2.3 class flare occurred on 2002
September 9 that were observed by the Reuven Ramaty High Energy
Solar Spectroscopic Imager (RHESSI) and the Owens Valley Solar Array
(OVSA). Interpreting the observed energy dependent variation of HXR
source size under the CTTM the coronal density should be as high as
$5\times 10^{11}$ cm$^{-3}$ over the distance up to 12$''$. To
explain the cut-off feature of microwave spectrum at 3 GHz, we
however, need density no higher than $1\times 10^{11}$ cm$^{-3}$.
Additional constraints need to be placed on temperature and magnetic
field of the coronal source in order to reproduce the microwave
spectrum as a whole. Firstly, a spectral feature called the Razin
suppression requires the magnetic field in a range of 250--350 gauss
along with high viewing angles around 75$^{\rm o}$. Secondly, to
avoid excess fluxes at high frequencies due to the free-free
emission that were not observed, we need a high temperature
$\geq$2$\times 10^7$ K. These two microwave spectral features,
Razzin suppression and free-free emissions, become more significant
at regions of high thermal plasma density and are essential for
validating and for determining additional parameters for the coronal
HXR sources.

\end{abstract}


\section{INTRODUCTION}
Observations of Hard X-ray (HXR) bursts with the Ramaty High Energy
Solar Spectroscopic Imager (RHESSI) have produced a number of new
properties of high energy electrons accelerated during solar flares.
One of its recent discoveries is a distinctive type of flares in
which the bulk of the HXR emission is from the coronal part of the
loop (Veronig \& Brown 2004; Sui et al. 2004; Krucker et al. 2008).
For such sources, the corona can be considered as not only the site
of particle acceleration, but act as a thick target, stopping the
accelerated electrons before they can penetrate to the chromosphere.
This is in contrast with the traditional idea of the collisional
thick target model (CTTM) that was believed to apply only to the
dense chromosphere. Veronig \& Brown (2004) studied two events for
which the coronal densities of the flaring loops were found
sufficiently high to justify the CTTM. Xu et al. (2008) showed that
for a certain group of events the observed coronal HXR source extent
varies with photon energy as expected from a CTTM with an extended
acceleration region in the coronal loop. The result not only serves
as evidence for the coronal CTTM, but also allows estimates of the
plasma density and longitudinal extent of the acceleration region.
Guo et al. (2012a) have carried out a similar work but using the
electron flux images instead of photon maps, as constructed by
regularized spectral inversion of the visibility data in the count
domain (Piana et al. 2007). Guo et al. (2012b) further related those
interpretations to the number of particles within the acceleration
region, the filling factor, and the specific acceleration rate in
units of electron per second and per ambient electron. According to
these studies those events with the coronal thick target HXR sources
form an important (though small) class of solar flares for which
acceleration parameters can directly be inferred.

If we plan to test the CTTM for coronal HXR sources using other
radiations, microwave radiation should be the first choice, for its
close relationship with HXRs. Time profiles of microwave emissions
tend to show a peak-to-peak correlation with HXR lightcurves, which
led to the idea of the same population of electrons responsible for
both radiations (e.g., Dulk 1985). Spatially both emissions should
arise, at least, from the same magnetic loop or loop systems, but
the location of each radiation maximum may differ along the loop as
their radiation efficiencies differently respond to plasma
parameters and magnetic field (e.g., Sakao 1996). We expect that
microwave spectrum will be particularly important in relation to the
studies of the coronal CTTM. The main microwave radiation mechanism,
gyrosynchrotron radiation, has primarily served as a diagnostic for
magnetic field and electron distribution. However, when thermal
density is very high, two spectral features sensitive to density,
the Razin suppression (Razin 1960) and free-free emission, become
prominent. The Razin suppression refers to less efficient radiation
in the presence of ambient medium compared with the radiation in
vacuum, and the Razin frequency below which the effect is
significant is given by $\nu_R\approx 20~n/B_{\bot}$ Hz where $n$ is
electron density in units of cm$^{-3}$ and $B_{\bot}$ is the
transverse magnetic field component in gauss. The Razin suppression
will thus be effective for the condition of thick target looptop HXR
sources. The well-known free-free opacity is proportional to
$n^2T^{-1/2}$, with $T$, the plasma temperature. In addition, there
is an {\it absolute} cut-off of microwave radiation at the plasma
frequency which is a sole function of plasma density (see, e.g.,
Dulk 1985, Melrose 1980). With these features that become prominent
at high density, a microwave spectrum should play a critical role in
validating the CTTM applied to coronal sources.

In this Letter we compare the properties of a coronal HXR source
with those of microwave source detected in an event which we
consider a candidate for the coronal thick target source. The event
is the 2002 September 9 flare occurred in NOAA AR 0105 and acquired
GOES soft X-ray class M2.3. This event was observed by the Reuven
Ramaty High-Energy Solar Spectroscopic Imager (RHESSI) and the Owens
Valley Solar Array (OVSA), respectively, and is one of the rare
events for which spatial and spectral information at both
wavelengths is available for this type of study.

\begin{figure}
\begin{center}
\epsscale{1.3} \plottwo{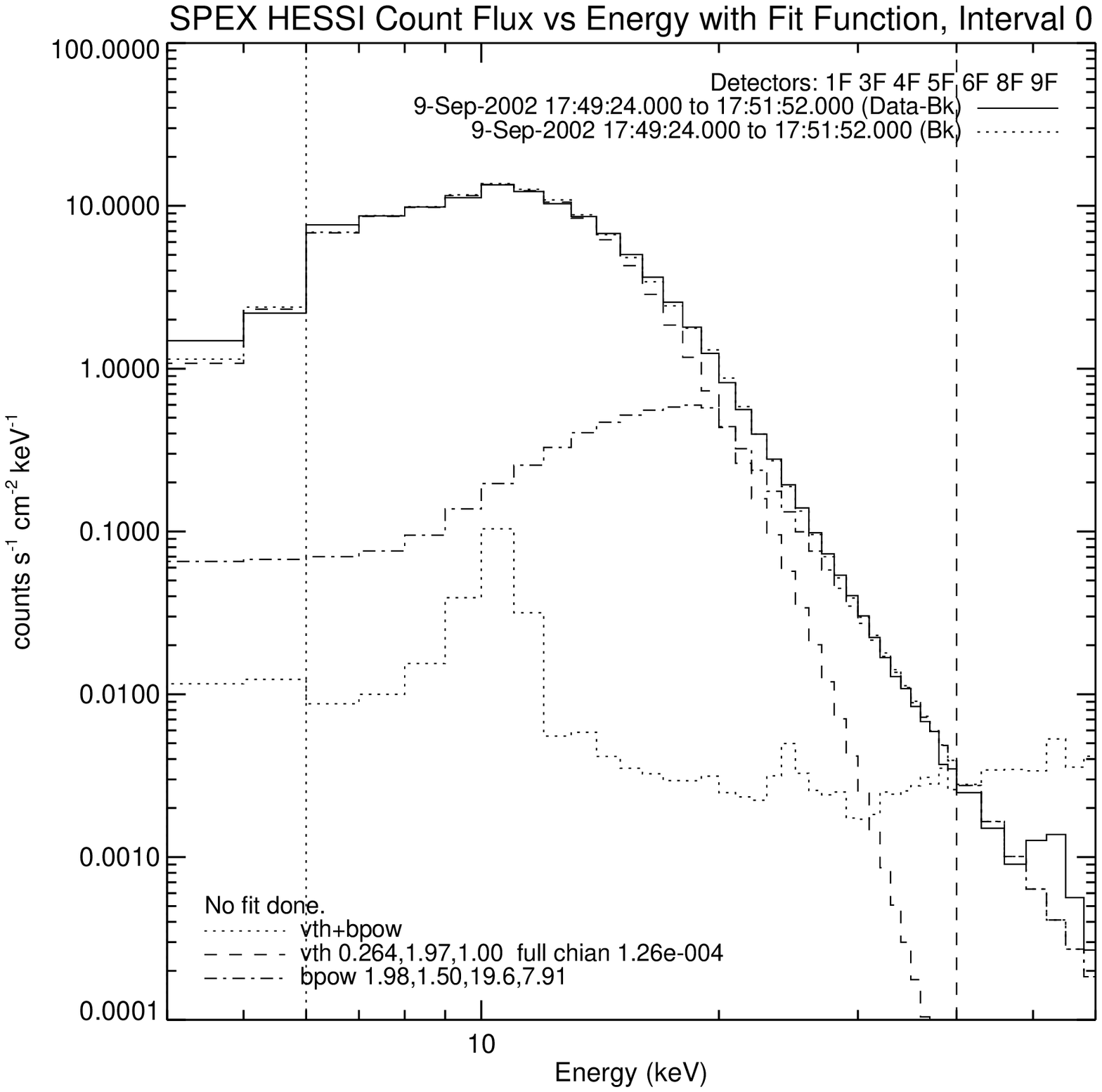}{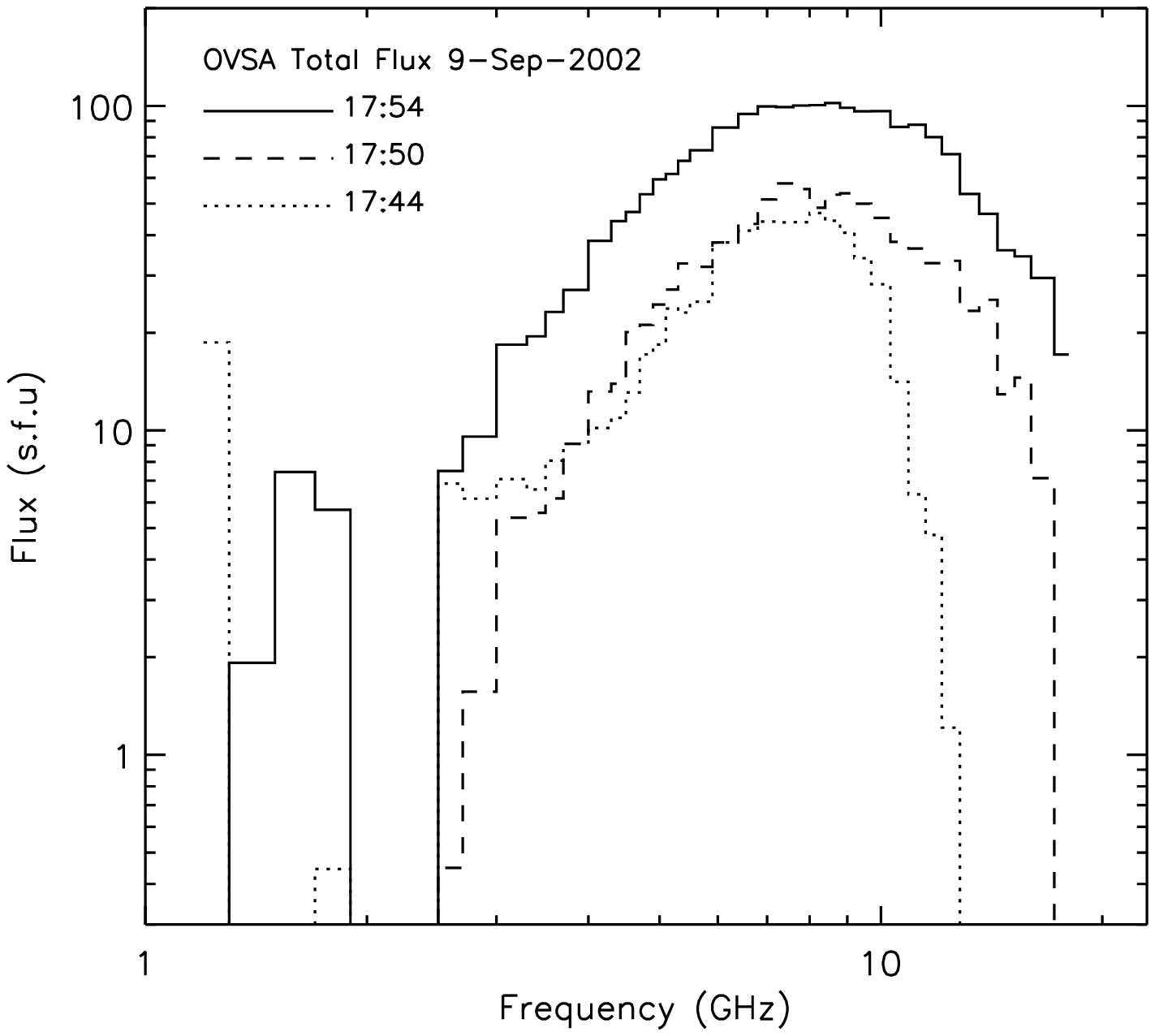} \caption{Upper panel: a
RHESSI spectrum for the flare on 2002 September 9. Spectral fit is
made for the interval, 17:49:23--17:51:53 UT. The thermal component
of the spectrum (dashed histogram) has temperature of 1.97 keV and
the nonthermal component (dot-dashed) has transition energy at 19.6
keV and spectral index $\gamma$= 7.91. The sum of the thermal and
nonthermal components (solid histogram) and the background is also
shown. Lower panel: the OVSA microwave spectra at three different
time intervals as denoted are shown.} \label{FIG01}
\end{center}
\end{figure}

\begin{figure}
\begin{center}
\epsscale{0.9} \plotone{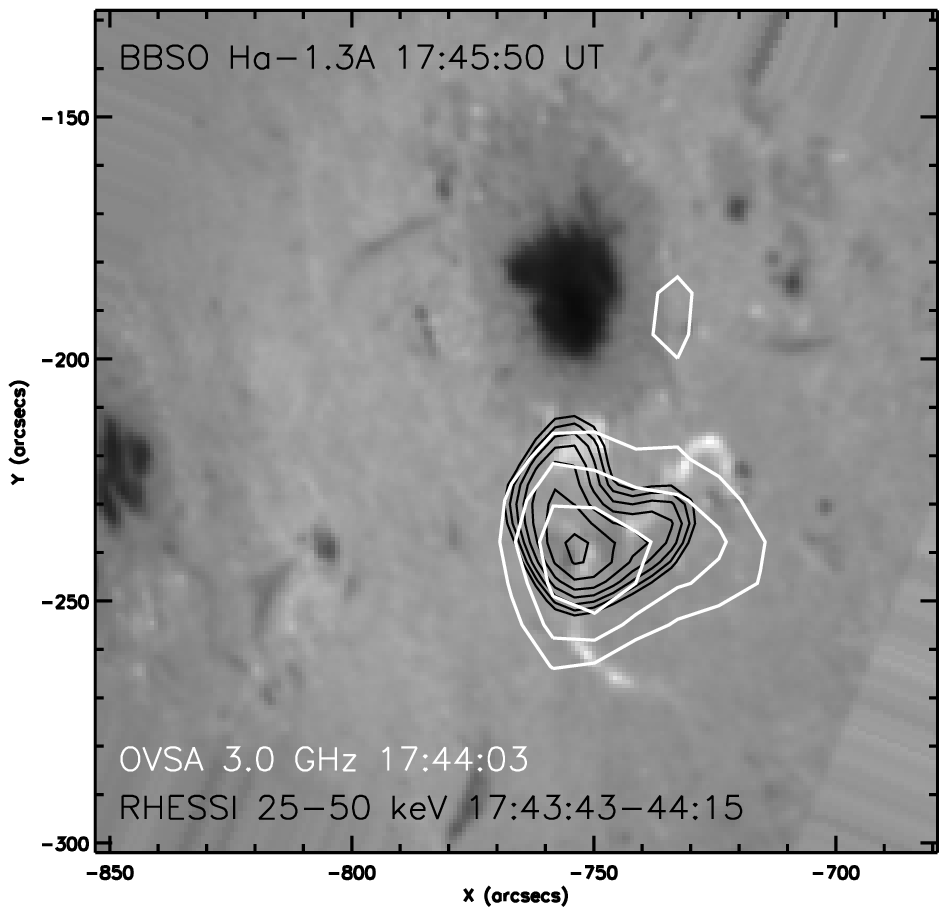} \caption{The 25-50 keV RHESSI source
(black contours) and the OVSA microwave source (white contours) in
comparison with the H$\alpha$ blue wing image (background greyscale
image) from the BBSO. These images are chosen to be near one of the
flare peak time 17:44 UT.} \label{FIG02}
\end{center}
\end{figure}

\section{SPECTRA OF HXR and Microwaves}

In Figure 1 we show the RHESSI and OVSA spectra during the 2002
September 9 flare. The top panel shows the spatially integrated
RHESSI spectrum at a single time corresponding to maximum emission
measure obtained using OSPEX in the SolarSoftWare (SSW).  Ji et al.
(2004) carried out the imaging spectroscopy for the same event. In
their result and also in ours (Fig. 2), the RHESSI source appears to
be a single source and the spectra obtained through OSPEX are indeed
similar to those obtained from the imaging spectroscopy. As a
characteristic feature, the spectrum shows steeply decreasing flux
with increasing photon energy with spectral indices as high as
7$\leq \gamma\leq$8. Such high index is typical for the events
classified for the coronal thick target HXR sources (Guo et al.
2012ab, Xu et al. 2007). The spectral index $\gamma$ will later be
used to provide the information on electron energy distribution to
the modeling of the HXR source loop and that of the microwave
spectra. In spite of such a soft nonthermal component, a thermal
component could still be identified and the maximum temperature is
found to be $T\approx 2\times 10^7$K at around 17:50 UT, close to
what we infer from GOES soft X-ray data.

The bottom panel shows the total power microwave spectra for the
2002 September 9 flare at three selected times. This is one of the
calibrated microwave spectra in the period of 2001-2003 that can be
downloaded from the OVSA web site.  Like the HXR spectra, these
microwave spectra are also spatially integrated and include both R
and L polarizations added together. They appear in a typical shape,
but with rapidly falling fluxes toward decreasing frequencies, which
can be a sign for the Razin suppression, and is suggestive of a high
coronal density. The high frequency spectrum is steeply decreasing
with increasing frequency, which is qualitatively consistent with
the high photon spectral index of the RHESSI spectrum as found
above. We may also note that the peak frequency of microwave
spectrum does not change much in the course of the flare.
Previously, Belkora (2007) claimed that such a feature may indicate
the effect of the Razin suppression. It is thus qualitatively
arguable that the spectral morphology of the OVSA spectra is
supportive of the high coronal density required for the CTTM of the
coronal HXR source.

\begin{figure}
\begin{center}
\epsscale{0.9} \plotone{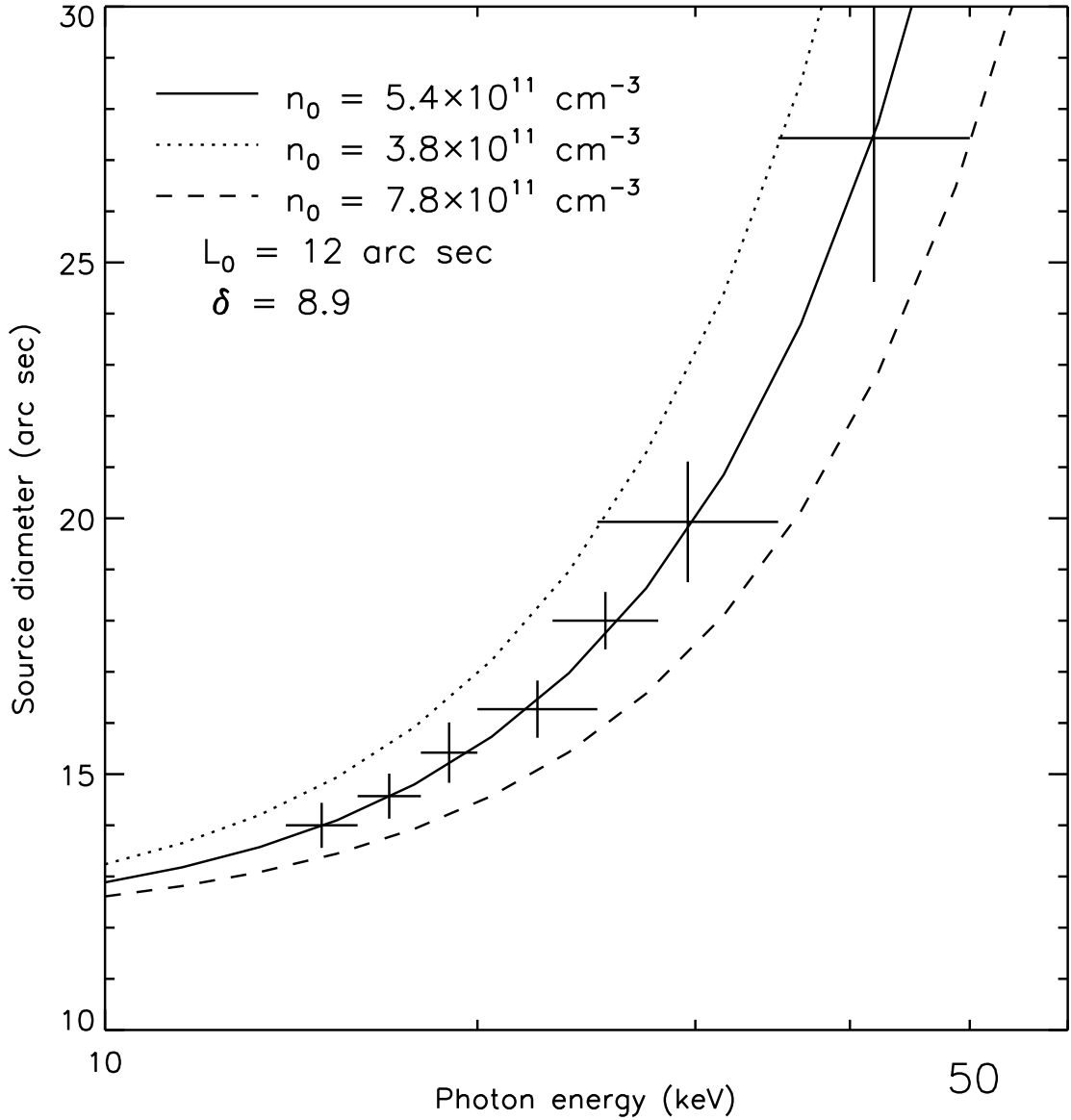} \caption{Model fits to the
longitudinal dimension of the HXR source. We calculate models
(lines) by varying ambient density $n_{\rm 0}$ and the initial
length of the trap $L_{\rm 0}$ to fit the observed HXR source
dimension (crosses) at seven photon energies.} \label{FIG03}
\end{center}
\end{figure}

\section{HXR and Microwave maps}

The OVSA imaging procedure is less known compared with that of the
RHESSI. Since the OVSA uses many frequencies, all night observations
of a quasar point source such as 3C84 are made several times a year,
and are used to calibrate solar data obtained at nearby times. Daily
calibration observations at selected frequencies are also made to
provide corrections to the above calibration. The calibrated
visibilities are fed into the {\it wimagr} program available in the
SSW, after which the imaging can be performed in the procedure very
similar to that of the Astronomical Image Processing System (AIPS).

Figure 2 shows HXR and microwave maps as contours on top of an
H$\alpha$ off-band blue-wing image obtained from the Big Bear Solar
Observatory (BBSO). Based on the steep HXR spectrum and the noise
level (Fig. 1), the HXR imaging was performed below 50 keV. Within
the energy range only single loop-top source was found. Microwave
sources generally vary with frequency more significantly, but the
centroid of 3.0 GHz microwave source is co-spatial with that of the
25--50 keV HXR source as shown in this figure. Both the HXR and
microwave sources appear as a single source lying between the two
H$\alpha$ kernels, and their morphologies change only a little with
time (see, for more details on the morphology of this event, Ji et
al. 2004 and Lee et al. 2006). We thus regard the HXR source as a
loop-top coronal thick target source (cf. Veronig \& Brown 2004). We
followed the technique of Xu et al. (2007) to determine the source
sizes using the Vis-Forward fitting and made a model fit to the
observed HXR source size as a function of photon energy. As for the
model, we actually used equation (5) of Guo et al. (2012b), which
corresponds to Xu et al.'s (2007) model with an extended
acceleration region of length, $L_{\rm 0}$, uniformly filled with
dense material of density $n_{\rm 0}$. This model also includes the
electron energy distribution with a power law index for which we use
$\delta=\gamma+1=8.9$ under the thick target bremsstrahlung
approximation (Brown 1971). Figure 3 shows our fitting of the model
to the observed HXR source size (symbols) at seven photon energies
by varying $n_0$ and $L_0$. We obtained $n_0 = (5.4\pm 1.6)\times
10^{11}$ cm$^{-3}$ and $L_0\approx 12''$ from these data.

\begin{figure}
\begin{center}
\epsscale{0.7} \plotone{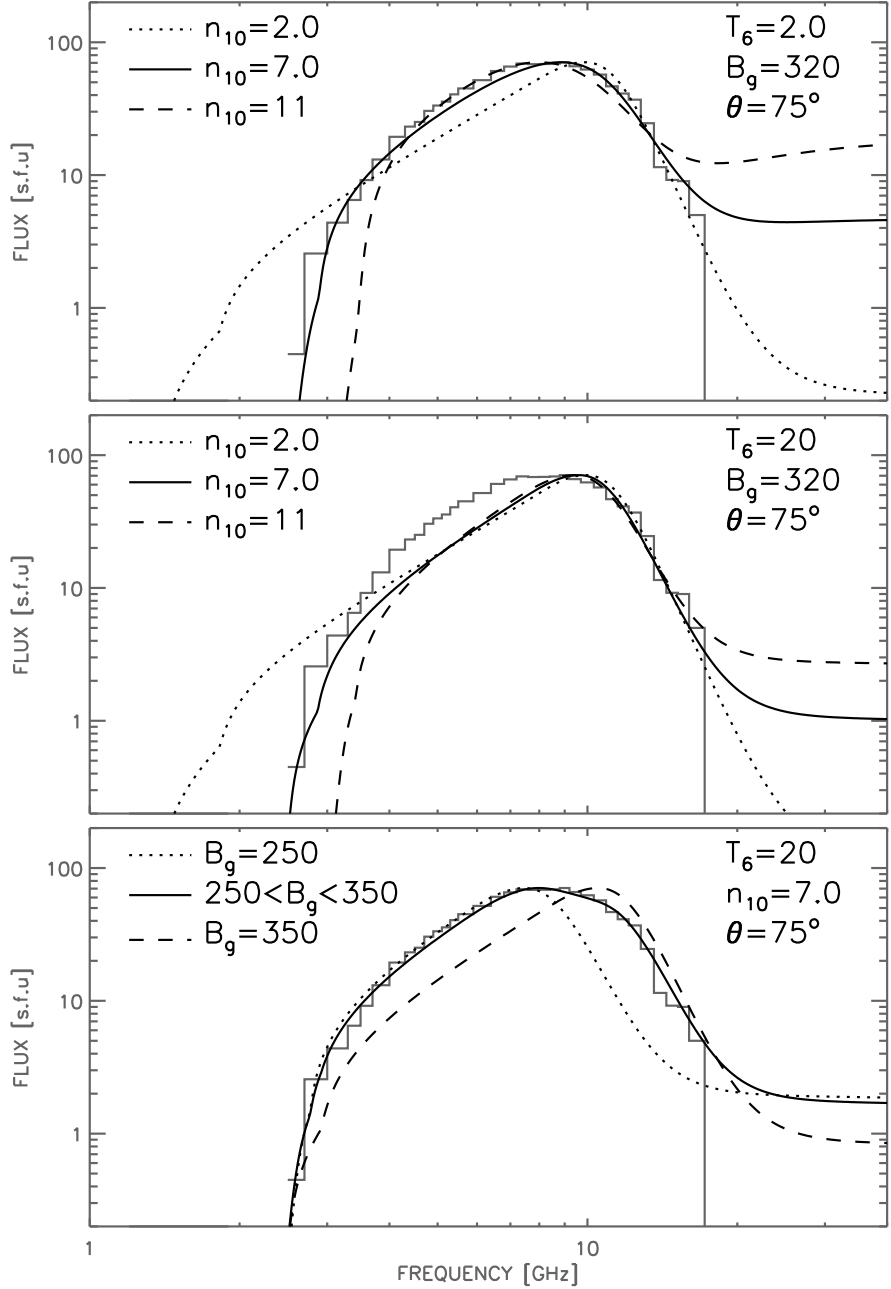} \caption{Model fits to an OVSA
spectrum observed at 17:50:30 UT. The OVSA spectrum is shown as a
histogram. Model parameters denoted are: n$_{\rm 10}$ (density in
units of $10^{10}$ cm$^{-3}$), T$_6$ (temperature in $10^6$ K), and
B$_{\rm g}$ (magnetic field in gauss). Top panel: three densities
are adopted to show how the spectrum varies with density. Middle
panel: the same model parameters as in the top panel are used except
that $T$ is raised to 10 times higher. Bottom panel: two models with
a weaker (stronger) magnetic fields (dotted and dashed lines) are
averaged to give an inhomogeneous model (solid line).} \label{FIG04}
\end{center}
\end{figure}

\section{Model Fits to the OVSA Spectra}

We calculated a set of theoretical microwave spectra for comparison
with the OVSA spectrum using the numerical code developed by
Fleishman and Kuznetsov (2010), which includes full treatment of the
Razin effect. In Figure 4 we show our search for the best set of
model parameters with which we can reproduce the OVSA spectrum at
the time of the maximum density ($\sim$17:50 UT) chosen as a
representative spectrum. In the top panel we first tried a modest
density $2.1\times 10^{10}$ cm$^{-3}$ and the same electron energy
distribution derived from the RHESSI spectrum, i.e., $\delta=8.9$,
except that the electron energy distribution is assumed to extend up
to 1 MeV. This model (shown as the dotted line in the top panel)
reproduces the high frequency spectrum to some extent, but fails to
match the low frequency spectrum. If a homogeneous model predicted a
more rapidly increasing spectrum at low frequencies than observed,
we may suspect that the disagreement is due to source inhomogeneity
in reality (e.g., Lee et al. 1994). In this case, however, the model
predicts a more extended spectrum toward lower frequencies than
observed. We thus adjusted other factors maintaining the homogeneous
source assumption. With a higher densities $7\times 10^{10}$
cm$^{-3}$ the model (solid line) reproduces the low frequency slope
under the enhanced Razin suppression. It closely fits to the high
frequency spectrum as well, except somewhat excessive fluxes are
predicted at the highest observed frequencies. If we further raise
the density as desired for supporting the thick target HXR source,
we end up with a disagreement between the model (dashed line) and
observation at both low and high frequencies, unfortunately.

At this point we should clarify one basic limitation encountered in
the current modeling. Even though we will further adjust other
parameters to reproduce the microwave spectrum more closely, the
density requirement for this microwave spectrum is rather strict.
That is, the minimum cut-off frequency for microwave radiation is
the local plasma frequency, a sole function of density. More
specifically, the o-mode is cut-off at local plasma frequency and
x-mode, at a higher frequency that depends on magnetic field as well
(Melrose 1980). Since this spectrum starts above 3 GHz, we derive
$n\leq 1\times 10^{11}$ cm$^{-3}$. Admittedly this density is a few
factor lower than the one needed for explaining the HXR source
dimension varying with energy under the CTTM, and this limit value
of density cannot be adjusted by change of other parameters.

We now proceed with the `lowered' target value of density. Since the
excessive fluxes at high frequencies are due to free-free emission,
we adopted higher temperatures as way of suppressing the free-free
emission. The three models shown in the middle panel inherit the
same parameters from those in the top panel except the temperature
increased to $2\times 10^7$ K. At this high temperature, the high
frequency fluxes are significantly {\it reduced}, and all models
agree to the observed spectrum at high frequencies. We may further
increase temperature without loosing this success, but too a high
temperature is unwanted in a physical point of view, because the
medium should be a cold-target for Coulomb collisions. Emslie (2003)
quantified the condition for cold target as $E > 5 kT$ where $E$ is
the electron energy and $T$ is the target temperature. Kosugi et al.
(1988) have shown that the electrons producing photons at $\epsilon$
have an energy in the range, $1 < E/\epsilon < 3$. Since we concern
ourselves with the RHESSI spectrum above 6 keV, the temperature
$2\times 10^7$ K or 2 keV may qualify as cold target. However, $T$
cannot be much higher than this, especially considering that this
photon spectrum is very soft and $E$ could be close to $\epsilon$.
For instance, $E/\epsilon<2 $ gives $T < 2.4\times 10^7$ K for cold
target.

In the above, change of the temperature helped to fit the high
frequency spectrum, but now these models no longer fit the low
frequency spectrum. To come up with this problem, we adjust the
magnetic field, $B$. In general, increasing (decreasing) $B$ results
in a shift of the spectrum to higher (lower) frequencies. Two models
calculated with $B=250$ gauss and 350 gauss are respectively shown
as the dotted and dashed lines in the bottom panel. Each individual
model spectrum appears to be narrower than the observed spectrum.
The low (high) field strength model could, however, reproduce the
low (high) frequency spectrum. We thus created an average model
(solid line) of these two  by linearly summing the flux from each
model, which well agrees to the observed spectrum in overall. We
thus conclude that the source must have an inhomogeneous magnetic
field distribution in the range of 250 gauss $\leq B \leq$ 350
gauss. The viewing angle of magnetic field, $\theta$, is another
variable and this should be high. Otherwise, the free-free emission
tends to compete with the gyrosynchrotron radiation, which is
unwanted in this case. An extensive exercise shows that a high
viewing angle is preferred, and we have fixed $\theta$ to 75$^{\rm
o}$ throughout this modeling. All these calculations were made with
the isotropic pitch angle distribution of electrons. If it were
significantly anisotropic, the resulting spectral morphology may
differ from the present result even for the same electron energy
distribution. In that case, a few more assumptions regarding the
electron energy distribution will be needed, which we do not regard
essential for the current study.

\section{Concluding Remarks}

We have studied the HXR and microwave data of the 2002 September 9
flare with a goal of testing the coronal thick target hypothesis for
HXR sources based on the accompanying microwave spectra. Our study
found that the microwave spectrum provides strong constraints on the
thick target interpretation of HXR sources. First of all, the lowest
frequency of the microwave spectrum is strictly determined by the
local plasma density, and the derived density $\leq 1\times 10^{11}$
cm$^{-3}$ is close to, but 5 times lower than the density found from
the energy dependent variation of the coronal HXR source. Not only
the density but temperature $\geq$2 keV and magnetic field strength,
250--350 gauss, are also required which will again be constrained by
the condition for a cold-target for bremsstrahlung and the
measurement of the photospheric magnetic field, respectively.

The next question will be whether these constraints could be
alleviated by introducing a more realistic, inhomogeneous density
structure instead of the simplified assumption for a uniform density
source. We commonly consider a source of dense and tenuous regions
mixed, i.e., filamentary structure, in which case the tenuous
regions filled with nonthermal electrons work as microwave sources
and the dense regions, the thick target HXR sources. This scenario
is not necessarily promising because microwaves should propagate
through the dense plasma and cannot avoid the highly enhanced
free-free opacity. The next conceivable structure is an
inhomogeneous loop along its length such that a section in loop-top
is tenuous, and the rest is filled with dense materials working as a
thick target to Coulomb collisions (Xu et al. 2007). This model is
actually favorable for explaining the microwave source, but has to
be disputed because it is in apparent conflict with the HXR source
lying in the looptop. In our current idea, the only plausible case
would be a somewhat arbitrary coaxial loop structure which consists
of the inner loop filled with dense colder plasma and the outer one
with tenuous and hot plasma, so that the former is ideal for the
thick target HXR source and the latter, more favorable for the
microwave emission. In this case the small disagreement between the
densities derived from microwaves and HXRs, respectively, could be
made acceptable. This scenario would, however, need an explanation
as to why the inner core has dense and colder materials compared
with the outer part of the loop.

In summary, microwave spectrum provides unique constraints on the
coronal HXR sources through the cut-off frequency, the cold target
temperature, the free-free opacity, and the Razin effect. These
constraints are related to the fundamental physics and are thus
robust. Microwave spectral observations should therefore be regarded
as a new asset for validating those events of the coronal HXR
sources and for determining additional source conditions for them
that were not available from HXR observations alone. Future studies
may be benefitted from the use of microwave polarizations and time
profiles that will further enhance its diagnostic capabilities.

\acknowledgments We thank \textit{RHESSI} team for excellent data
set. This work was supported by WCU Program through NRF funded by
MEST of Korea (R31-10016). J.L. was supported by the International
Scholarship of Kyung Hee University, and also by NASA grants,
NNX11AB49G and NAS5-01072 (NA03) 937836 to NJIT. G.S.C. was
supported by the Korea Research Foundation grant funded by the
Korean Government (KRF-2007-313-C00324). M. J. expresses his thanks
to Kyung Hee University for granting him a sabbatical, during which
his part of this research was performed.

\newpage


\begin{references}

\reference{ref01} Belkora, L. 1997, \apj, 481, 532

\reference{ref01} Brown, J. C. 1971, \solphys, 18, 489

\reference{ref01} Dulk, G. A. 1985, ARA\&A, 23, 169

\reference{ref01} Emslie, A. G. 2003, \apj, 595, L119

\reference{ref01} Fleishman, G. D. \& Kuznetsov, A. A. 2010, \apj,
721, 1127

\reference{ref01} Guo, J., Emslie, A. G., Kontar, E. P., et al.
2012a, A\&A, 543, 53

\reference{ref01} Guo, J., Emslie, A. G., Massone, A. M., and Piana,
M., 2012b, \apj 755, 32

\reference{ref01} Ji, H., Wang, H., Schmahl, E. J., Qiu, J., Zhang,
Y. 2004, \apj, 605, 938

\reference{ref01} Kosugi, T., Dennis, B. R., \& Kai, K, 1988, \apj,
324, 1118

\reference{ref01} Krucker, S., Hurford, G. J., MacKinnon, A. L.,
Shih, A. Y., \& Lin, R. P. 2008, \apj, 678, 63

\reference{ref01} Lee, J. \& Gary, D. E. 2000, \apj, 543, 457

\reference{ref01} Lee, J., Gary, D. E., \& Choe, G. S., 2006, \apj,
647, 638

\reference{ref01} Lee, J., Gary, D. E., \& Zirin, H. 1994, \solphys,
152, 409

\reference{ref01} Piana, M., Massone, A. M., Hurford, G. J., et al.
2007, \apj, 665, 846

\reference{ref01} Ramaty, R. 1969, \apj, 158, 753

\reference{ref01} Razin, V. A. 1960, Izv. VUZov. Radiofiz., 3, 584

\reference{ref01} Sakao, T., Kosugi, T., Masuda, S., Yaji, K.,
Inda-Koide, M., \& Makishima, K. 1996, AdSpR, 17, 67

\reference{ref01} Sui, L., Holman, G. D., \& Dennis, B. R. 2004,
\apj, 612, 546

\reference{ref01} Melrose, D. B. 1980, Plasma Astrophysics, Gordon
and Breach, Science Publishers, Inc.

\reference{ref01} Veronig, A., and Brown, J. 2004, \apj, 603, L117

\reference{ref01} Xu, Y., Emslie, A. G., \& Hurford, G. J. 2008,
\apj, 673, 576



\end{references}
\end{document}